\documentclass[technote]{IEEEtran}
\title{\bf Reducing the  Computation of Linear Complexities of Periodic 
 Sequences over $GF(p^m)$}
\author{Hao Chen
  \thanks{H. Chen is with the Department of Computing and 
          Information Technology, School of Information Science
          and Engineering, Fudan University, Shanghai, 200433,
          People's Republic of China}}

\begin{document}

\maketitle
\begin{abstract}
The linear complexity of a periodic sequence over $GF(p^m)$ plays an important role in cryptography and %%@
communication [12]. In this correspondence, we prove a result which
reduces the
 computation of the linear complexity and minimal connection polynomial of a period $un$
sequence  over $GF(p^m)$ to the computation of the linear
complexities and minimal connection polynomials of $u$
 period $n$ sequences. The conditions $u|p^m-1$ and
 $\gcd(n,p^m-1)=1$ are required for the result to hold. Some applications of this %%@
reduction in fast algorithms to determine the linear complexities
and minimal connection
polynomials of sequences over $GF(p^m)$ are presented.
\end{abstract}


\section{Introduction}
For a period $N$ sequence ${\bf a}=a_0,a_1,...,a_{N-1},a_0,...$ over a finite field $GF(p^m)$,  its linear %%@
complexity $c({\bf a})$ is defined to be the length of the shortest linear feedback shift register to generate it, i.e. the %%@
smallest positive integer $k$ such that there exist some $c_1,...,c_k$ in $GF(p^m)$ and %%@
$a_{i+k}=c_1a_{i+k-1}+ \cdots +c_ka_i$ hold for all  $i \geq 0$. The polynomial $m({\bf a})=1-(c_1x+ \cdots +c_kx^k)$  is called the minimal connection polynomial [12].\\

The linear complexity of a periodic sequence is considered as the measure of its randomness and plays an important role %%@
in the application of the sequence in cryptography and
communication. There are many works [1],[2],[4],[6],[8]
,[9],[10],[11],[14],[15] and [16] on efficient algorithms %%@
for determining the linear complexities and minimal connection polynomials of sequences. Some authors also %%@
have interesting results about the linear complexities of some special sequences (see [3],[7] and [13]).  The famous %%@
Berlekamp-Massey algorithm [11] can be used to compute the linear complexity and minimal connection polynomial of a  %%@
period $N$ sequence over $GF(p^m)$ with time complexity $O(N^2)$ (that is, at most $O(N^2)$ field operations in %%@
$GF(p^m)$).  One of the main advantages of the Berlekamp-Massey
algorithm is the input at the step $t$ of the algorithm is the first
$t$ elements of the sequence. Actually, the Berlekamp-Massey
algorithm only needs $2c({\bf a})$  consecutive elements of the
sequence to determine its linear complexity and minimal connection
polynomial [12]. An adapted fast version of Berlekamp-Massey
algorithm due to Blackburn [1] can be used with
time complexity $O(N(\log N)^2\log\log N)$.\\

In [6] Games and Chan  gave a fast algorithm which can be used to determine the linear complexity and minimal connection %%@
polynomial of a period $N=2^t$ binary sequence  with time complexity $O(N)$. This algorithm was also generalized %%@
to compute the linear complexity and minimal connection polynomial of a period $N=p^t$ sequence over $GF(p^m)$  with %%@
time complexity $O(N)$ (see [5] and [8]). Based on the Games-Chan
algorithm, some authors developed fast algorithms
 [9], [10] and [14] for computing the k-error linear complexities of  period $N=2^t$ binary sequences
 and period $N=p^t$ sequences over $GF(p^m)$. G.Xiao et al. [15] and [16] gave fast algorithms to compute the linear complexities and minimal connection %%@
polynomials of period $N=p^t$ or $N=2p^t$ sequences over $GF(q)$ , when $q$ is a primitive root modulo %%@
$p^2$. For sequences of period $N=2^tn$, where $2^t|p^m-1$ and
$\gcd(n,p^m-1)=1$, a fast algorithm which
can be used to determine their linear complexities %%@
 more efficiently was given in our paper [4].\\

It is well known that the linear complexity and  minimal connection
polynomial of a periodic sequence
over $GF(p^m)$ can be %%@
understood from its generating function. For a sequence ${\bf
a}=a_0,a_1,...,a_{N-1},a_0,...$ over $GF(p^m)$ of period $N$,
its generating function $A(x)=a_0+a_1x+ \cdots +a_ix^i+ \cdots =\Sigma_{i \geq 0}a_ix^i=\frac %%@
{a_0+a_1x+ \cdots +a_{N-1}x^{N-1}}{1-x^N}$. Then the linear complexity of the sequence ${\bf a}$ %%@
is $c({\bf a})=\deg(1-x^N)-\deg(\gcd(a_0+a_1x+ \cdots +a_{N-1}x^{N-1},1-x^N))$ and the minimal connection polynomial is %%@
$m({\bf a})(x)=\frac {1-x^N}{\gcd(a_0+a_1x+ \cdots +a_{N-1}x^{N-1},1-x^N)}$ [12].\\

In this correspondence we prove a result which reduces  the
computation of the linear complexity
 and  minimal connection %%@
polynomial of a period $un$ sequence over $GF(p^m)$ to the
computation of the linear complexities and minimal connection
polynomials of $u$
 period $n$ sequences. This reduction result can be combined with various known %%@
algorithms to compute the linear complexities of sequences more
efficiently. The main result of this correspondence can be thought
as a generalization
 of the result in our previous paper [4].

\section{Main Result}
Let $m$ be a positive integer, $p$ be a prime number, $u$ be a positive integer such %%@
that $u$ divides $p^m-1$, and $n$ be a positive integer such that
$\gcd(n,p^m-1)=1$ . It is clear there are $u$ distinct $u$-th roots
of unity  $x_0,...,x_{u-1}$, where $x_0=1$, in $GF(p^m)$ since
$u|p^m-1$. From the condition  $\gcd(n,p^m-1)=1$, we can find a
unique $b_i \in GF(p^m)$, which is the $n$-th root of $x_i$ for all
$i=0,...,u-1$. The following result is the main result of this correspondence.\\

{\bf Theorem.}  {\em   Suppose $p, m, u, n, x_0,...,x_{u-1},
b_0,...,b_{u-1}$ are given as above. Let ${\bf
a}=a_0,a_1,...,a_{un-1},a_0,a_1,...$ be a  period $un$ sequence over
$GF(p^m)$. Let ${\bf a^j}$ be the period $n$ sequence over $GF(p^m)$
 with its first period
$a_0+a_nb_j^{n}+ \cdots
+a_{(u-1)n}b_j^{(u-1)n},...,a_ib_j^{i}+a_{n+i}b_j^{n+i}+ \cdots
+a_{(u-1)n+i}b_j^{(u-1)n+i},...,
a_{n-1}b_j^{n-1}+a_{2n-1}b_j^{2n-1}+ \cdots +a_{un-1}b_j^{un-1}$,
for $j=0,1,...,u-1$.  Then $c({\bf a})=c({\bf a^0})+c({\bf a^1})+
\cdots+c({\bf a^{u-1}})$ and $m({\bf a})(x)=m({\bf
a^0})(b_0^{-1}x)m({\bf a^1})(b_1^{-1}x) \cdots m({\bf a^{u-1}})(b_{u-1}^{-1}x)$.}\\

{\bf Proof.} Let $f(x)=\Sigma_{i=0}^{un-1}a_ix^i$. It is clear $
1-x^{un}= \prod_{i=0}^{u-1} (x_i-x^n)=x_1 \cdots x_{u-1} (1-x^n)
\prod_{i=1}^{u-1} (1-(b_i^{-1}x)^n)$. Any two distinct polynomials
among the $u$ polynomials $(1-x^n)$, $(1-(b_1^{-1}x)^n)$,...,
$(1-(b_{u-1}^{-1}x)^n)$ are coprime in
$GF(p^m)[x]$. Thus $\gcd(f(x),1-x^{un})=\gcd(f(x),1-x^n) \prod_{i=1}^{u-1} \gcd(f(x),1-(b_i^{-1}x)^n)$.\\

It is clear $\gcd(f(x),1-x^n)=\gcd(f_0(x),1-x^n)$, where
$f_0(x)=\Sigma_{i=0}^{n-1}(a_i+a_{n+i}+ \cdots +a_{(u-1)n+i})x^i$.
Thus $c({\bf a^0})=\deg(\frac{1-x^n}{\gcd(f(x),1-x^n)})$ and $m({\bf
a^0})(x)=\frac{1-x^n}{\gcd(f(x),1-x^n)}$. For each $j$ satisfying $1
\leq j \leq u-1$, we set $\gcd(f(x),1-(b_j^{-1}x)^n)=g_j(x)$ and
$\gcd(f(b_j y),1-y^n)=h_j(y)$. Then $g_j(x)=h_j(b_j^{-1}x)$.  We
have $h_j(y)=\gcd(f_j(y),1-y^n)$, where
$f_j(y)=\Sigma_{i=0}^{n-1}(a_ib_j^{i}+a_{n+i}b_j^{n+i}+ \cdots
+a_{(u-1)n+i}b_j^{(u-1)n+i})y^i$. Thus $c({\bf
a^j})=\deg(\frac{1-y^n}{h_j(y)})$ and $m({\bf
a^j})(y)=\frac{1-y^n}{h_j(y)}$. Finally $c({\bf
a})=un-[\Sigma_{i=0}^{u-1} \deg(\gcd(f(x),1-(b_{i}^{-1}x)^n))]
=c({\bf a^0})+c({\bf a^1})+ \cdots +c({\bf a^{u-1}})$ and $ m({\bf
a})(x)=m({\bf a^0})(b_0^{-1}x)m({\bf a^1})(b_1^{-1}x)
\cdots m({\bf a^{u-1}})(b_{u-1}^{-1}x)$. The conclusion is proved.\\

When $u=2^tn$, the above result was proved in our
previous paper [4].

In the reduction we need the storage of $u$ elements $b_0=1,b_1,...,b_{u-1} \in GF(p^m)$  in advance.
For a period $N=un$ sequence  over $GF(p^m)$, where $u|p^m-1$ and $\gcd(n,p^m-1)=1$,  we need  %%@
$\frac{(u-1)N}{u}$ field operations to get the sequence ${\bf a^0}$, $(u-1)N$ field operations to get the elements \\
$ b_1,...,b_1^{(u-1)n-1},...,b_{u-1},...,b_{u-1}^{(u-1)n-1}$,  and $\frac{(2u-1)(u-1)N}{u}$ field %%@
operations to get the sequences ${\bf a^1},...,{\bf a^{u-1}}$. Thus the time complexity of the reduction in the main result %%@
is $3(u-1)N$ field operations in $GF(p^m)$.\\

\section{Applications}
In this section we use the main result and some known algorithms to
give fast algorithms
for computing the linear complexities of sequences over $GF(p^m)$.

{\bf A. An easy example}\\
Let $p$ be an odd prime, $m$ be an arbitrary  positive integer and $n$ be a positive integer such that $n$ and $p^m-1$ %%@
are coprime. Then we have a unique element $b$ in $GF(p^m)$ such
that $b^n=-1$. Here we note $b^{2n}=1$. For arbitrary
$a_0,...,a_{n-1} \in GF(p^m)$, let ${\bf
a}=a_0,a_1,...,a_{n-1},-a_0,-a_1,...,-a_{n-1},a_0,...$ be a period
$2n$ sequence over $GF(p^m)$ . Set  ${\bf
a'}=2a_0,2a_1b,...,2a_ib^i,...,2a_{n-1}b^{n-1},2a_0,...$ , which is
a period $n$ sequence over $GF(p^m)$ .
From the main result, the linear complexity $c({\bf a})$ is the same as the linear %%@
complexity $c({\bf a'})$ and the minimal connection polynomial
$m({\bf a})(x)$ is just $m({\bf a'})(bx)$. Thus the linear
complexity and
 minimal connection polynomial of
the period $2n$ sequence ${\bf a}$ can be determined from the period $n$ sequence ${\bf a'}$.\\

{\bf B. Combining with the  generalized Games-Chan algorithm}\\
In this subsection it is  assumed that $p$ is a prime number, $m$ is
a positive integer and $u$ is a positive integer such that $u$
divides $p^m-1$. We now give a fast algorithm to compute the linear
complexity $c({\bf a})$ of a period $N=up^h$  sequence ${\bf a}$
over $GF(p^m)$ with time complexity $O(N)$. Here $u$ is understood
as a constant not depending on the
sequence. We need the storage of $u$ elements $b_0=1, b_1,...,b_{u-1}$  in advance.\\

{\bf Input:} A period $N=up^h$ sequence ${\bf a}$  over $GF(p^m)$.\\

{\bf Output:} The linear complexity $c({\bf a})$.\\

{\bf Algorithm.}\\

Perform the reduction of the main result, we get $u$ period $p^h$ sequences ${\bf a^0},...,{\bf a^{u-1}}$.\\

For the period $p^h$ sequences ${\bf a^0},...,{\bf a^{u-1}}$,
perform the following generalized Games-Chan algorithm
{\bf GGC}, the outputs are the linear complexities $c({\bf a^0}),...,c({\bf a^{u-1}})$.\\

{\bf GGC Algorithm.}\\

1) Initial value: ${\bf s}  \leftarrow{\bf s}=(s_0,...,s_{p^h-1})
\in
GF(p^m)^{p^h}$, $N \leftarrow p^h$, $c \leftarrow 0$.\\

2) Repeat the following a)-c) until $h=0$.\\

a) For a given $p^h$-tuple ${\bf s}$, set ${\bf
s^{(i)}}=(s_{ip^{h-1}},...,s_{ip^{h-1}+p^{h-1}-1})$ for
$i=0,...,p-1$,  and ${\bf b^{(u)}}=\Sigma_{j=0}^{p-u-1}C_{p-j-1}^u
{\bf s^{(j)}}$,  where $u=0,...,p-1$ and $C_{p-u-1}^u$'s are the
binomial coefficients.\\

b) Find the smallest $w$ such that  ${\bf b^{(0)}}={\bf
b^{(1)}}=...={\bf b^{(p-w-1)}}=0$ and ${\bf b^{(p-w)}} \neq 0$ for a
$w \in \{1,...,p\}$. Here if ${\bf b^{(0)}} \neq 0$, we set $w=p$.\\

c) Do ${\bf s} \leftarrow {\bf b^{(p-w)}}$, $c \leftarrow (w-1)p^{h-1}+c$, and goto a).\\

3) When $h=0$ and ${\bf s}=(s_0) \neq  0$, then $c \leftarrow c+1$, otherwise $c \leftarrow c$.\\

The final output $c$ of ${\bf GGC}$ is the linear complexity $c({\bf
s})$ of the period $p^h$ sequence ${\bf s}$  over $GF(p^m)$.\\

Finally we get the linear complexity of $c({\bf a})=\Sigma _{i=0}^{u-1} c({\bf a^{i}})$ from the main result.\\

We refer to [5],[8] and [10] for the generalized Games-Chan
algorithm. {\bf GGC} needs at most $2p^2N'$
 field operations in $GF(p^m)$ for determining the linear complexity of a period $N'=p^h$ sequence over
 $GF(p^m)$.
 On the other hand we need at most $3(u-1)N$ field operations in the reduction
  for a given period $N=uN'$ sequence. Thus the above algorithm
 needs $3(u-1)N+u(2p^2\frac{N}{u})=[3(u-1)+2p^2]N$ field operations in $GF(p^m)$, where $N$ is the period of
 the input sequence. The coefficient $3(u-1)+2p^2$ is a fixed constant not depending on the
 sequence. For example,
 the above fast algorithm can be used to determine the linear complexities of
  period $N=3 \cdot 7^h$ sequences
 over $GF(7^m)$
 and period $N=3 \cdot 13^h$ sequences  over $GF(13^m)$.\\

{\bf Example.} Let ${\bf a}=123401520113061256331....$ be a period
$21$ sequence over $GF(7)$. We want to compute its linear complexity
and minimal connection polynomial
by the above algorithm. First we note $b_0=1$, $b_1=4$ and $b_2=2$ in $GF(7)$. Then\\

${\bf a^0}=4424645, {\bf a^1}=4366203, {\bf a^2}=2622130$.\\

$c({\bf a})=c({\bf a^0})+c({\bf a^1})+c({\bf a^2})$.\\

$m({\bf a})(x)=m({\bf a^0})(x)m({\bf a^1})(4x)m({\bf a^2})(2x)$.\\

In the case of $p=7$ we use the generalized Games-Chan algorithm
and get\\

$c({\bf a^0})=7,m({\bf a^0})=(1-x)^7$,\\

$c({\bf a_1})=7,m({\bf a^1})=(1-x)^7$,\\

$c({\bf a^2})=7,m({\bf a^2})=(1-x)^7$.\\

Finally we have $c({\bf a})=21$ and $m({\bf a})=(1-x)^7(1-4x)^7(1-2x)^7$.\\

Comparing with the Blackburn's algorithm given in [2], the reduction
to the $u$ period $p^h$ sequences is the same as that in the
Blackburn's algorithm, because in this case the $u$-th root of unity
$\alpha$ in
[2] is an element of $GF(p^m)$.\\

{\bf C. Combining with the Berlekamp-Massey algorithm }\\

We can also apply the reduction of the main result to compute the
linear complexity of a period $N=un$ ($\gcd(n,p^m-1)=1$) sequence
${\bf a}$ over $GF(p^m)$, where $u$ divides $p^m-1$ and $n$ is not a power of $p$. In this %%@
case, we apply the Berlekamp-Massey algorithm [11] with time complexity $O(n^2)$ (or the Blackburn's version [1] of %%@
Berlekamp-Massey algorithm with time complexity $O(n(\log n)^2\log
\log n)$ ) to the $u$  period $n$ sequences after the reduction. It
is obvious that this would be more efficient than applying the
Berlekamp-Massey algorithm directly to the original sequence.
However when this reduction is used, we have to
know the whole period of the sequence.\\

{\bf D. Combining with the  Xiao-Wei-Lam-Imamura algorithm}\\

Let $p$ and $q$ be two prime numbers. Suppose  $q$ is a primitive root modulo $p^2$, that is, $q$ is the generator %%@
of the multiplicative group of residue classes (modulo $p^2$) which
are coprime to $p$, then a fast algorithm for determining the linear complexity of a period $N=p^n$ sequence over $GF(q^m)$ with time complexity %%@
$O(N)$ was given in [16]. Combining with the reduction in our main
result, we can determine the linear complexity
of a  period $N=up^n$ sequence  over $GF(q^m)$ with time complexity $O(N)$, if $u$ divides $q^m-1$ , $q$ is a primitive %%@
root modulo $p^2$,  $p$ and $q^m-1$ are coprime. For example, it is
easy to check that $13$ is a primitive root modulo
 $25$, thus we can determine the linear complexities of period $N=3 \cdot 5^n$ sequences over $GF(13^m)$(if $m \neq %%@
0$, $mod$ $ 4$) with time complexity $O(N)$.\\

{\bf IV. Conclusion}\\
We have proved a result reducing the computation of the linear
complexity of a period $un$ sequence  over $GF(p^m)$, where $u$
divides $p^m-1$ and $\gcd(n,p^m-1)=1$,  to the computation of the
 linear complexities of $u$ period $n$
sequences . Based on this reduction and some known algorithms we can
compute the linear complexities
 of period $un$ sequences  over $GF(p^m)$  more efficiently. It seems that the main result might be useful for other %%@
problems about the linear complexities of sequences over $GF(p^m)$.\\

{\bf Acknowledgment.} The author wishes to thank the Associate
Editor and  the anonymous referees for their helpful comments and
criticisms on the 1st version of the paper. This work was supported
in part by   NNSF of China under  Grant 90607005 and
Distinguished Young Scholar Grant 10225106.\\

e-mail: chenhao@fudan.edu.cn\\

\begin{center}
REFERENCES
\end{center}

[1] S.R.Blackburn, Fast rational interpolation,Reed-Solomon decoding and the linear complexity profiles of sequences, %%@
IEEE Trans. Inf. Theory, vol.43, no.2, pp.537-548, Mar. 1997.\\

[2] S.R.Blackburn, A generalization of the Discrete Fourier
Transform: determining the minimal polynomial of a periodic sequnce,
IEEE Trans. Inf. Theory, vol.40, no.5, pp.1702-1704, Sept. 1994.\\

[3] S.R.Blackburn, The linear complexity of the self-shrinking generator, IEEE Trans. Inf. Theory,vol.43, no.5,%%@
 pp.2073-2077, Sept. 1999.\\

[4] H.Chen, Fast algorithms for determining the linear complexity of
sequences over $GF(p^m)$ with period $2^tn$, IEEE Trans. Inf. Theory, vol.51, no.5, pp.1854-1856, May 2005.\\

[5] C.Ding, G.Xiao and W.Shan, The stability theory of stream ciphers, Lecture Notes in Computer Science, vol.56, %%@
Springer-Verlag, 1991.\\

[6] R.A.Games and A.H. Chan, A fast algorithm for determining the complexity of a binary sequence with period $2^n$, %%@
IEEE Trans. Inf. Theory, vol.29, no.1, pp.144-146, Jan. 1983.\\

[7] F. Griffin and I.E.Shparlinski, On the linear complexity profile of the power generator. IEEE Trans. Inf. %%@
Theory,  vol.46, no.5, pp.2159-2162, Sept.2000. \\

[8] K.Imamura and T.Moriuchi, A fast algorithm for determining the linear complexity of p-ary sequences with period %%@
$p^n$, p prime, IEICE Tech. Rep. IT 93-75, pp.73-78, 1993.\\

[9] T.Kaida, S.Uehara and K.Imamura, An algorithm for the k-error linear complexity of sequences over $GF(p^m)$ with %%@
period $p^n$, p prime, Information and Computation, vol. 151, pp.134-147, 1999.\\

[10] A.G.Lauder and K.G.Paterson, Computation the error linear complexity spectrum of a binary sequence with period %%@
$2^n$, IEEE Trans. Inf. Theory, vol.49, no.1, pp.273-280, Jan.2003.\\

[11] J.L.Massey, Shift register synthesis and BCH decoding, IEEE Trans. Inf. Theory, vol.15, no.1, pp.122-127, Jan.1969.\\

[12] A. Menezes, P. van Oorschot and S.Vanstone, Handbook of applied cryptography, CRC Press Inc. 1997.\\

[13] I.E.Shparlinski, On the linear complexity of the power generator, Designs, Codes and Cryptography, 2001, vol.23, no.1, %%@
pp.5-10, Jan.2001. \\

[14] M.Stamp and C.F.Martin, An algorithm for the k-error linear complexity of binary sequences with period $2^n$, IEEE %%@
Trans. Inf. Theory, vol.39, no.4, pp.1398-1401, July 1998.\\

[15] S.Wei, G.Xiao and Z.Chen, A fast algorithm for determining the minimal polynomial of a sequence with period %%@
$2p^n$ over $GF(q)$, IEEE Trans. Inf. Theory, vol.48, no.10, pp2754-2758, Oct.2002.\\

[16] G.Xiao, S.Wei, K.Y.Lam and K.Imamura, A fast algorithm for determining the linear complexity of a sequence with %%@
period $p^n$ over $GF(q)$, IEEE Trans. Inf. Theory, vol.46, no.6, pp.2203-2206, Sept.2000.\\

{\bf Hao Chen} was born in Anhui Province of China on Dec. 21, 1964. He is currently a professor in the Department of %%@
Computing and Information Technology of Fudan University, Shanghai,
China. His research interests are cryptography and coding, quantum
information and computation.

\end{document}